# SEEPAGE FLOW-STABILITY ANALYSIS OF THE RIVERBANK OF SAIGON RIVER DUE TO RIVER WATER LEVEL FLUCTUATION

A. Oya[1], H.H. Bui[2], N. Hiraoka[1], M. Fujimoto[3], R. Fukagawa[3]

[1]Graduate school of Science and Engineering, Ritsumeikan University, Japan,

[2]Department of Civil Engineering, Monash University, Australia  [3]Department of Civil Engineering, Ritsumeikan University, Japan

**ABSTRACT:** The Saigon River, which flows through the center of Ho Chi Minh City, is of critical importance for the development of the city as forms as the main water supply and drainage channel for the city. In recent years, riverbank erosion and failures have become more frequent along the Saigon River, causing flooding and damage to infrastructures near the river. A field investigation and numerical study has been undertaken by our research group to identify factors affecting the riverbank failure. In this paper, field investigation results obtained from multiple investigation points on the Saigon River are presented, followed by a comprehensive coupled finite element analysis of riverbank stability when subjected to river water level fluctuations. The river water level fluctuation has been identified as one of the main factors affecting the riverbank failure, i.e. removal of the balancing hydraulic forces acting on the riverbank during water drawdown.

## 1. INTRODUCTION

Many cities in south-east Asian countries have developed in downstream areas of the regions great rivers, and as a result they often suffer from flooding. Ho Chi Minh City (HCMC), located in southern Vietnam, is one of the leading economic and commercial hubs in South-East Asia. The Saigon River runs through the center of HCMC, provides its main source of water, and contributes to its industrial development. However, riverbank failure has recently become a serious issue especially at the flood events, with numerous reports of settlement and in some cases buildings collapsing. An example of a riverbank failure in the Thanh Da peninsula region (along the Saigon River) is shown in Figure 1. Furthermore, the increasing frequency of riverbank failures and subsequent flooding may hinder the future economic development of HCMC [1].

Examples of riverbank failure countermeasures currently in use along the Saigon River are shown in Figure 2. These include: wooden piles to structurally reinforce river banks, and soil bags as temporary reinforcement measures. However, reinforcement of large sections of the river with piles is likely to be prohibitively expensive, sandbags are at best of a practical short term solution; an effective and economical countermeasure is required. However, in order to identify effective countermeasures it is necessary to first develop an understanding of the mechanisms leading to riverbank instability and failure and their causes. Investigating riverbank failure mechanisms has been a focus of our research group. The river water level fluctuation has been identified as one of the main factors affecting the riverbank failure [2]. However, the seepage behavior due to water fluctuations and its effect on riverbank stability have not been revealed. In this paper, field investigation results obtained on the Saigon River are presented, followed by a coupled finite element analysis of riverbank stability when subjected to river water level fluctuation.

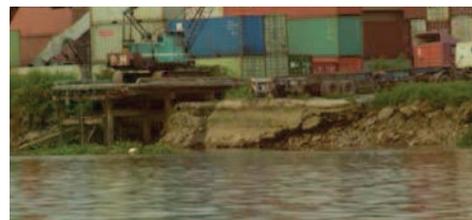
Figure 1 Riverbank failure along Saigon River

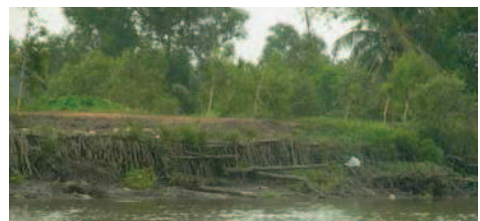
Figure 2 Countermeasure by wood piles

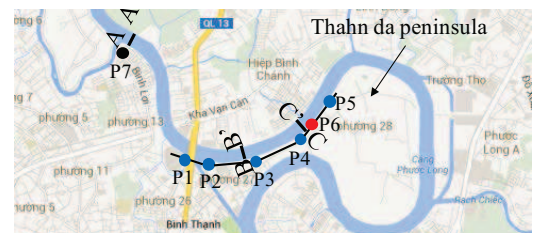
Figure 3 Locations of the observation sites

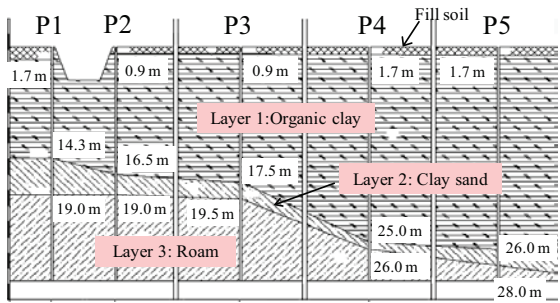

Figure 4 Soil layers at P1-P5

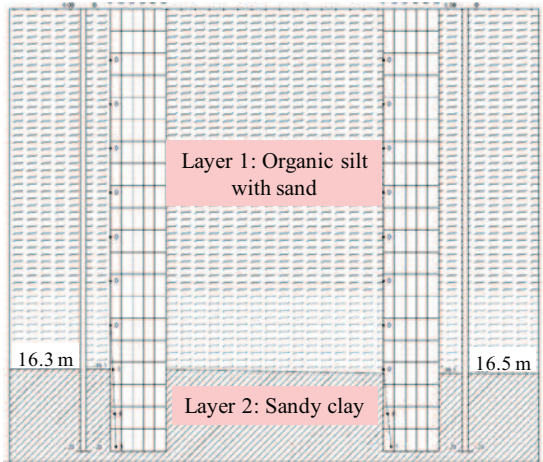

Figure 5 Soil layers at P6

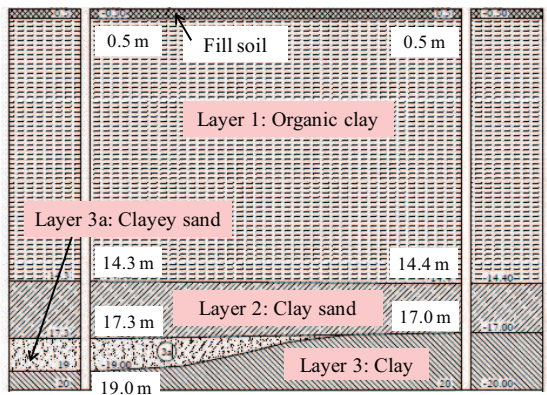

Figure 6 Soil layers at P7

## 2. GEOTECHNICAL INVESTIGATION

Site investigations have been conducted at seven observation points along the banks of the Saigon River, most of which occur around the Thahn Da peninsula in HCMC (Figure 3). Standard Penetration Tests (SPT) and comprehensive laboratory soil testing (Moisture content, Wet density, Specific gravity, Atterberg limits, Triaxial test or Direct shear strength test and consolidation test) based on ASTM criteria were performed on soil samples recovered from sites P1–P5 (blue dots), 2 points at site P6 (red dot), and site P7 (black dot). Bathymetry surveys of the Saigon River were undertaken using an Acoustic Doppler Current Profiler (ADCP) at 3 locations (Figure 3). The results of ADCP testing are presented in Figures 7 to 9. Also, groundwater level and river water level were measured at site P7 (Figure 3).

### 2.1 Sub-surface conditions

Three cross-sectional profiles have been determined at investigation site P1 to P5, P6 and P7, and cross sections are presented in Figures 4, 5 and 6 respectively. Soil parameters obtained from the soil testing are presented Tables 1, 2 and 3.

Table 1 Soil parameters at P1-P5

| Item | | Layer1 | Layer2 | Layer3 |
|---|---|---|---|---|
| $w_L$ | (%) | 70 | 36 | |
| $w_P$ | (%) | 38 | 18 | |
| $I_p$ | (%) | 32 | 18 | |
| $w$ | (%) | 75 | 20 | 17 |
| $\rho_t$ | (g/cm³) | 1.47 | 2.11 | 2.11 |
| $\rho_d$ | (g/cm³) | 0.86 | 1.76 | 1.81 |
| $Gs$ | | 2.61 | 2.72 | 2.66 |
| $Sr$ | (%) | 95.2 | 99.5 | 96.6 |
| $c$ | (kN/m²) | 10.8 | 36.3 | 9.8 |
| $\phi$ | (deg) | 4.38 | 15.7 | 28.1 |
| $k$ | (m/sec) | 6.2×10⁻⁸ | 4.8×10⁻⁷ | 6.1×10⁻⁵ |
| $N$ | | 0-3 | 8-15 | 15-30 |

Table 2 Soil parameters at P6

| Item | | Layer1 | Layer2 |
|---|---|---|---|
| $w_L$ | (%) | 87.2 | 55.6 |
| $w_P$ | (%) | 44.9 | 27.0 |
| $I_p$ | (%) | 42.3 | 28.6 |
| $w$ | (%) | 95.7 | 46.8 |
| $\rho_t$ | (g/cm³) | 1.46 | 1.76 |
| $\rho_d$ | (g/cm³) | 0.74 | 1.19 |
| $Gs$ | | 2.62 | 2.70 |
| $Sr$ | (%) | 97.0 | 96.9 |
| $c'$ | (kN/m²) | 11.8 | 16.6 |
| $\phi'$ | (deg) | 18.9 | 23.2 |
| $k$ | (m/sec) | 4.02×10⁻¹⁰ | 1.62×10⁻¹⁰ |
| $N$ | | 0 | 3-7 |

Table 3 Soil parameters at P7

| Item | | Layer1 | Layer2 | Layer3 | Layer3a |
|---|---|---|---|---|---|
| $w_L$ | (%) | 73.4 | 39.2 | 51.1 | |
| $w_P$ | (%) | 46.8 | 20.3 | 30.2 | |
| $I_p$ | (%) | 26.6 | 18.9 | 20.9 | |
| $w$ | (%) | 86.7 | 26.6 | 44.0 | 27.0 |
| $\rho_t$ | (g/cm³) | 1.45 | 1.95 | 1.71 | 1.88 |
| $\rho_d$ | (g/cm³) | 0.77 | 1.52 | 1.19 | 1.48 |
| $Gs$ | | 2.60 | 2.72 | 2.72 | 2.67 |
| $Sr$ | (%) | 95.0 | 92.0 | 93.0 | 89.0 |
| $c$ | (kN/m²) | 11.3 | 23.7 | 17.2 | 17.2 |
| $\phi$ | (deg) | 9.43 | 12.3 | 23.4 | 7.55 |
| $k$ | (m/sec) | 5.95×10⁻⁶ | - | - | - |
| $N$ | | 1-2 | 12-13 | 7-11 | 8 |

Symbols used for the tables indicate as follows; $w_L$: Liquid limit, $w_p$: Plastic limit, $I_p$: plastic index, $w$: water content, $\rho_t$: Wet density, $\rho_d$: Dry density, $Gs$: Specific gravity of soil particle, $Sr$: Degree of saturation, $c$: cohesion, $\phi$: Friction angle, $k$: Hydraulic conductivity, $N$: SPT-N-value. At each of the investigation locations there were typically 2 or 3 distinct soil units. At P7, 4 soil layers were identified. At all locations, underlying the fill unit was a very soft organic clay/silt with a thickness varying from about 14 m to 26 m. The presence, and considerable thickness of this very soft soil, is of particular importance for slope stability assessment of the Saigon riverbank.

**2.2 Geometry of the bank**

ADCP was used to investigate the three-dimensional distribution of flow-velocities in the Saigon River by obtaining three cross sections at locations shown in Figure 3. ADCP measures the river flow-velocities by transmitting ultrasonic waves into the floodway, and the results of this investigation are shown in Figures 7 to 9. Riverbank slopes of 30° occur at both sites, and heavy and thin lines in the river channels indicate the riverbed and the range of measurement, respectively.

**2.3 River water levels and ground water levels**

Measurement of the groundwater level was carried out at three wells, numbered 1, 3, and 4; these are presented in Figure 10. Similarly, the river water level was measured using a water pressure gauge placed directly on the river bed.

Figure 11 shows variations in the groundwater level and river water level at observation site P7 over the duration of the experiment. A zero-value for the level of both river water and groundwater was set as equal to the ground level at 1. Groundwater level fluctuations reach a maximum of approximately 0.5 m, despite the low permeability of the soil. Larger fluctuations can be observed at observation points closer to the river, and the groundwater level is higher on at observation points furthest away. These data show that the Saigon River water level fluctuates by about 2 m in a single day, and the relationship between river water levels and groundwater levels.

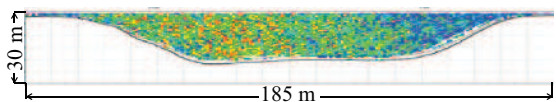
Figure 7 Cross-section at A-A' (ADCP No.1)

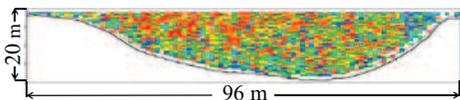
Figure 8 Cross-section at B-B' (ADCP No.2)

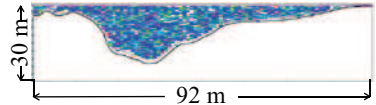
Figure 9 Cross-section at C-C' (ADCP No.3)

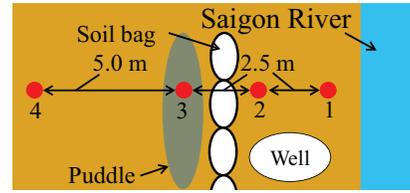
Figure 10 Set up condition of the wells

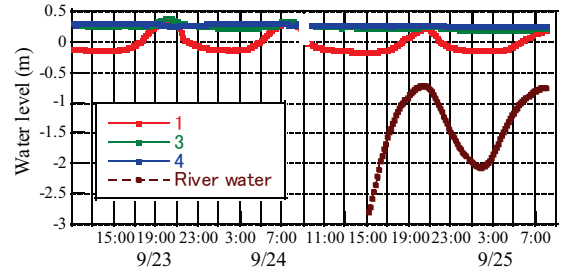
Figure 11 Groundwater level / River water level

**3. FLOW-STABILITY ANALYSIS**

Seepage flow-stability coupled analyses using commercial FEM software PLAXIS was conducted to examine how the water fluctuation and infiltration characteristic of the bank affect stability of the riverbank.

**3.1 Analyses outlines**

FEM analyses of the riverbank subjected to river water fluctuation were performed using PLAXIS [5]. Seepage flow-stability coupled analyses are undertaken. The analyses cover 3 points that site investigations are done. Soil properties at P7, P1-5 and P6 are applied to ADCP profiles No.1, No.2, and No.3 respectively. Soil properties at each site are combined with cross section at nearest points of each site for soil properties. Young modulus $E$ is correlated from N-value of the ground, Poisson's ratio $\nu$ is set at 0.33 which is typical value for geomaterials, and. Dilatancy angle $\psi$ is set at 0 [6]. Other parameters are determined by the results of laboratory soil tests mentioned above. Slope models are based on ADCP survey, and we develop the slope models by reading directly the line in the pictures. The slope model at No.1 is partly based on the field survey. Constitutive law is linear elastic perfectly plastic model (Mohr-Coulomb model). Shear strength reduction technique is applied to slope stability analyses. Unsaturated infiltration characteristic is determined by the Van-Genuchten model. Then, in order to assess seepage behavior in the bank, river water fluctuation is given as a boundary condition on river side of the bank. The observed water fluctuation of 2 m is assessed as part of this analysis.

### 3.2 Analyses conditions

*3.2.1 Conditions for No.1*: The geometry of the slope stability model is presented in Figure.12, analysis parameters are presented in Table 4. The model consists of 3 layers based on the subsurface conditions and soil parameters which were obtained from the site investigation. Soil permeability tests have not been conducted on Layer 2 and 3 materials at this stage thus hydraulic conductivity is assumed based on particle size distribution [7].

The initial water condition is based on an initial river water level and phreatic level set at 19.5 m from the bottom of the slope, this was determined by site survey. The water level is then linearly reduced to 17.5 m over a period of 4 hours, kept at 17.5 m for an hour, and then increased to 19.5 m over a period of 4 hours and kept at this level for an hour. Furthermore, initial phreatic line in the ground is assumed to be straight at 19.5m which is same as river water level.

*3.2.2 Conditions for No.2*: The geometry of the slope model is shown in Figure.13, and analysis parameters are shown in Table 5. The model consists of 3 layers based on the subsurface condition and soil parameters which were obtained from the site investigation. The initial river water level is set at 15.5m from the bottom of the slope, which is determined from the result of ADCP measurement. The initial phreatic line in the ground is assumed to be a straight line and the water level reduces 2 m over the period of 4 hours, kept at 13.5 m for an hour, increase to 15.5 m over the period of 4 hours and kept at this lever for an hour.

*3.2.3 Conditions for No.3*: Slope model is shown in Figure.14, and the analysis parameter is shown in Table 6. The slope model consists of 2 layers based on the results of the site investigation as well as other points. Initial river water level is set at 22.0 m, which is determined by result of ADCP measurement. initial phreatic line in the ground is assumed to be straight as well The water level take a descend of 2 m spending 4 hours as well as the point of profile A-A'. Then, it is kept at 13.5 m for an hour, takes an ascent of 2 m spending 4 hours and kept for an hour.

### 3.3 Analyses results and discussion

For the analysis for No.1 (profile A-A'), results are shown in Figure.15 as potential failure surface obtained by stability analyses. The blue lines show

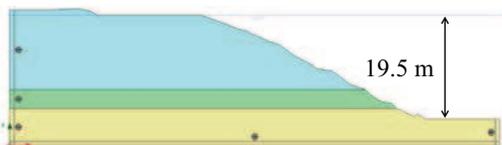

Figure 12 Slope model at A-A'

Table 4 Analysis parameter for A-A'

| Item | | Layer1 | Layer2 | Layer3 |
|---|---|---|---|---|
| $\gamma_t$ | (kN/m²) | 13.2 | 19.1 | 16.8 |
| $\gamma_{sat}$ | (kN/m²) | 14.1 | 19.2 | 17.6 |
| $E$ | (MPa) | 4.20 | 27.3 | 22.4 |
| $\nu$ | - | 0.33 | 0.33 | 0.33 |
| $c$ | (kN/m²) | 11.3 | 23.7 | 17.2 |
| $\phi$ | (deg) | 9.43 | 12.28 | 23.38 |
| $\psi$ | (deg) | 0 | 0 | 0 |
| $k$ | (m/s) | 5.95×10⁻⁶ | 3.00×10⁻⁸ | 1.00×10⁻⁸ |

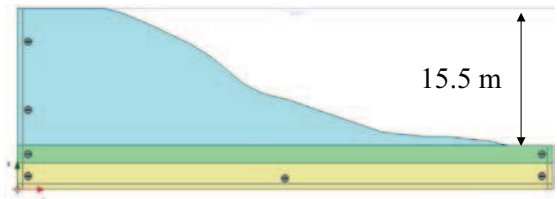

Figure 13 Slope model at B-B'

Table 5 Analysis parameter for B-B'

| Item | | Layer1 | Layer2 | Layer3 |
|---|---|---|---|---|
| $\gamma_t$ | (kN/m²) | 14.72 | - | - |
| $\gamma_{sat}$ | (kN/m²) | 15.2 | 20.7 | 21.1 |
| $E$ | (MPa) | 5.60 | 33.6 | 64.4 |
| $\nu$ | - | 0.33 | 0.33 | 0.33 |
| $c$ | (kN/m²) | 10.79 | 36.3 | 9.81 |
| $\phi$ | (deg) | 4.38 | 15.7 | 28.1 |
| $\psi$ | (deg) | 0 | 0 | 0 |
| $k$ | (m/s) | 6.2×10⁻⁸ | 4.8×10⁻⁷ | 6.1×10⁻⁵ |

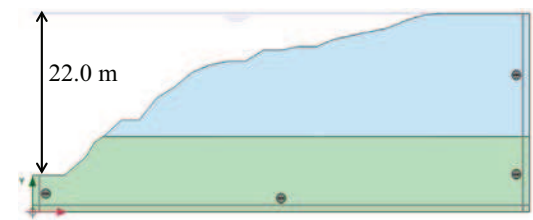

Figure 14 Slope model at C-C'

Table 6 Analysis parameter for C-C'

| Item | | Layer1 | Layer2 |
|---|---|---|---|
| $\gamma_t$ | (kN/m²) | 14.72 | - |
| $\gamma_{sat}$ | (kN/m²) | 15.19 | 20.74 |
| $E$ | (MPa) | 5.60 | 33.60 |
| $\nu$ | - | 0.33 | 0.33 |
| $c$ | (kN/m²) | 10.79 | 36.30 |
| $\phi$ | (deg) | 4.38 | 15.66 |
| $\psi$ | (deg) | 0 | 0 |
| $k$ | (m/s) | 6.2×10⁻⁸ | 4.8×10⁻⁷ |

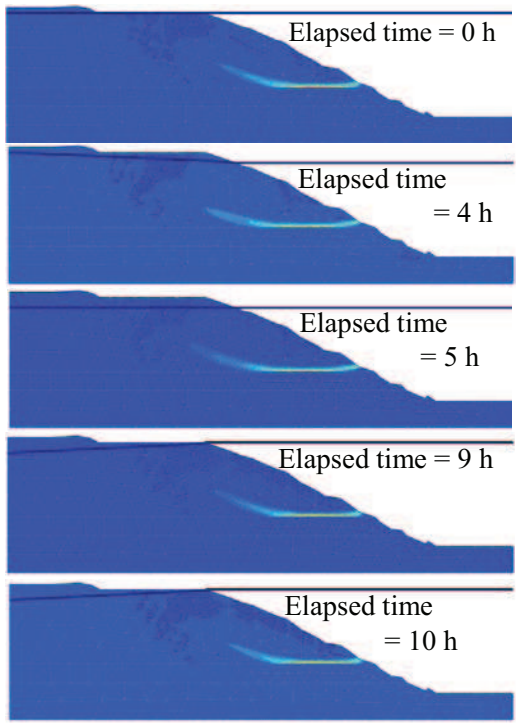

Figure 15 Potential failure surface on profile A-A'

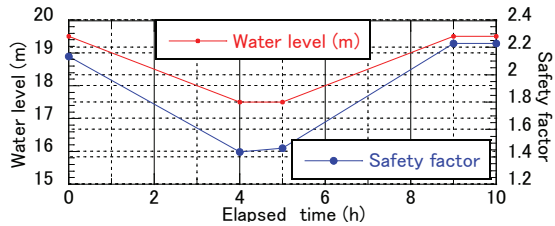

Figure 16 Safety factor on profile A-A'

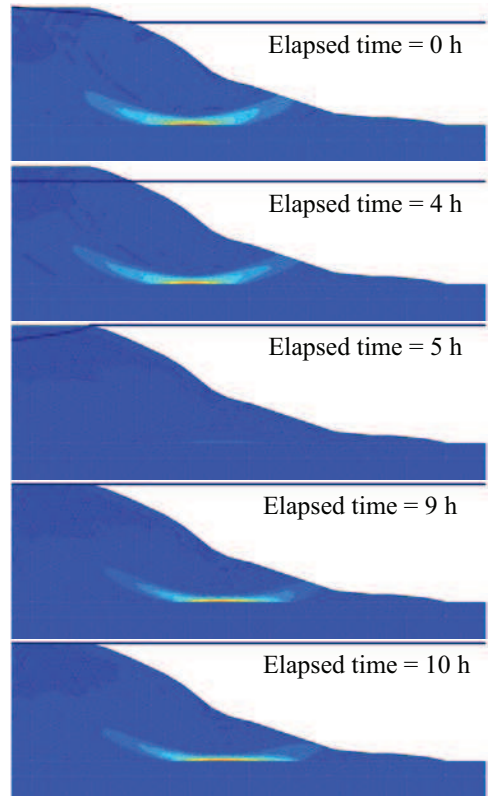

Figure 17 Potential failure surface on profile B-B'

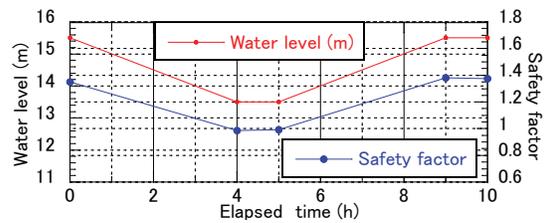

Figure 18 Safety factor on profile B-B'

the estimated the phreatic level in the ground. Safety factors are plotted in Figure 16 with river water level. When the water level goes down, the safety factor is lower. The riverbank slope is stable because of hydraulic pressure from the river when the water level remains high; however, when the river water level goes down, the hydraulic pressure dissipates and the slope becomes unstable. When the phreatic level in the river water level goes down, the phreatic level in the ground remains at a high level because of the ground's low permeability, and the weight of the soil remains high. It can be assumed that this leads to the instability of the slope. Furthermore, potential failure surface occurs in deeper point in lower water level stage, which indicates more massive slope failure can occur in this situation.

As for the analysis for No.2, results are shown in Figure 17 as potential failure surface obtained by stability analyses. The blue lines show the estimated the phreatic level in the ground. Safety factors are plotted in Figure 18 with river water level. At 5h, the water level ascent period, significant slip surface does not appear. In other stage of analyses, failure surfaces are observed on the boundary between layer 1 and 2. A safety factor value under 1.00, which indicates the slope is unstable, is estimated in the period of water level descent. The maximum safety factor value of the slope at B-B' is lower than any safety factors of the slope at A-A' and C-C'. From the aforementioned factors, the slope is originally unstable, and gets more unstable due to water level descent.

The results of the analysis for No.3 (profile C-C') are shown in Figure 19 as potential failure surface obtained by stability analyses. The safety factors of the bank estimated stability analyses are s are plotted in Figure 20. The safety factor of the bank falls when the water level decreases and remains low, as is the case with profile A-A'. The phreatic level remains higher than the river water level, which is assumed to be the cause of the instability of the slope. The difference in safety factors for water level decent, the low water level period, water level ascent and the high water level period is smaller than that of other points analyzed. This is because the height of the slope is bigger than other points. As for slip surfaces, significant ones do not appear when a water level is constant such as elapsed time 0 h and 5 h. On the other hand, 2 failure surfaces on the boundary of layer 1 and 2 and toe of the slope can be observed at elapsed time 4 h. There is a possibility that most massive slope failure occurs in this period.

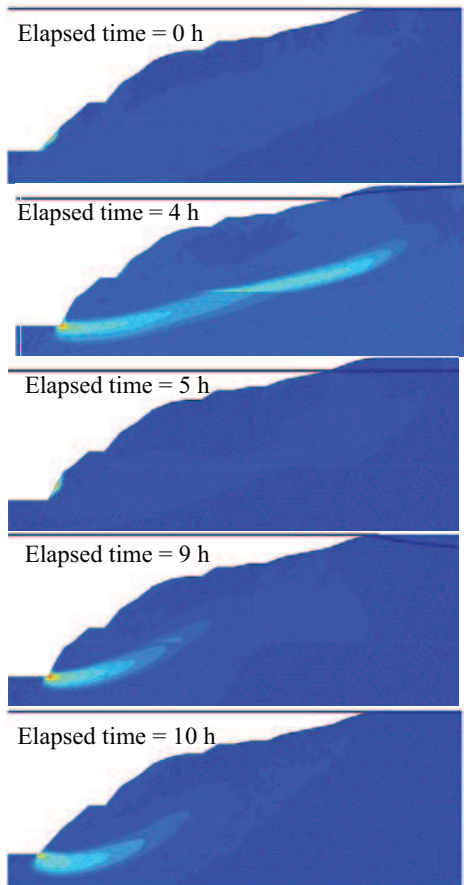

Figure 19 Potential failure surface on profile C-C'

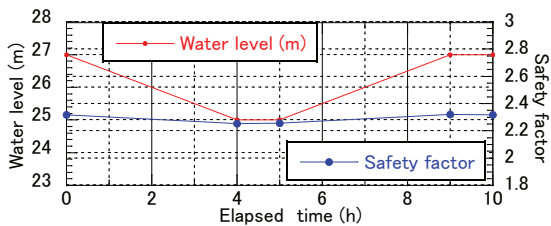

Figure 20 Safety factor on profile C-C'

## 4. CONCLUSION

In this paper, geotechnical properties and hydraulic conditions along the Saigon River have been presented, and seepage flow-stability analysis was conducted. Soft clay soil is deposited as thick horizons, and the ground is mechanically weak at the all of points where we have carried out our investigation. As results of our analysis, we can observed that the safety factors of the slopes are degraded during water level descent and the low water level period, which is a typical phenomenon that we can observe in the analyses of rapid drawdown of reservoir level in the dam[8]. These results suggest that river water fluctuation affects the stability of a riverbank, especially during water level descent because the phreatic level in the river water level goes down, the phreatic level in the ground remains at a high level because of the ground's low permeability, and the weight of the soil remains high, the slope is stable when water level is higher on the other hand. For future investigation, slope stability and seepage behavior due to cyclic water level fluctuations should be examined.

## 5. ACKNOWLEDGEMENTS

This research was financially supported by Grants-in-Aid for scientific research (B) 23404012. Also, authors would thank Dr. LUU Xuan Loc (Ho Chi Minh City University of Technology) for his help in carrying the site investigations.